\documentclass[cameraready]{Interspeech}
\usepackage{makecell}
\usepackage{algorithm}
\usepackage{algorithmic}
\usepackage{multirow}
\usepackage{makecell}

\title{An Analysis of the Effectiveness of Synthetic Speech Data for ASR Fine-tuning in Selected Indic Languages}

\author[affiliation={1}]{Sujith}{Pulikodan}
\author[affiliation={1}]{Agneedh }{Basu}
\author[affiliation={1}]{Pavan}{Kumar}
\author[affiliation={1}]{Pranav}{Bhat}
\author[affiliation={1}]{Visruth}{Sanka}
\author[affiliation={1}]{Nihar}{Desai}
\author[affiliation={2}]{Prasanta }{Kumar Ghosh}

\address{
    $^1$ AI \& Robotics Technology Park (ARTPARK), I-Hub @ IISc, Bangalore, India \\
    $^2$ Department of Electrical Engineering, Indian Institute of Science, Bangalore, India 
}

\email{sujith@artpark.in}


\usepackage{comment}

\begin{document}

\maketitle

\begin{abstract}
Synthetic data has the potential to be a valuable resource for training machine learning models, particularly Automatic Speech Recognition (ASR) Systems; however, its effectiveness requires systematic evaluation. In this study, we investigate the impact of incorporating synthetic speech data alongside real-world recordings for three Indic languages: Hindi, Kannada, and Telugu. We analyze the performance gains achieved by augmenting synthetic data with real data and independently examine how ASR performance varies with the sources of scripts used to generate synthetic speech. In addition, we evaluate the effect of synthetic speech generated using different speech synthesis models. Finally, we study the impact of voice cloning in synthetic speech generation on ASR performance, including how performance varies with the number of distinct cloned voices used during data generation.
\end{abstract}

\section{Introduction}

The process of speech data collection is both expensive and time-consuming \cite{hughes2010building}\cite{novotney2010cheap}. Developing robust machine learning models, especially for speech-related tasks, requires access to large and diverse datasets. However, for many Indic languages, the availability of such data remains a significant challenge \cite{joshi-etal-2020-state}. Although several initiatives have been undertaken to collect and annotate speech data across Indian languages, these efforts often result in limited datasets that are insufficient for training large-scale models. Factors such as the linguistic diversity of India, the presence of numerous dialects, and the lack of standardized resources further complicate the data collection process \cite{census2011language}.

To address this scarcity, synthetic data generation has emerged as a promising solution for data augmentation. Synthetic data aims to fill the gap between the vast data requirements of modern machine learning systems and the limited availability of real-world data. The use of generative models—such as text-to-speech (TTS) systems, voice conversion models, and diffusion-based speech generators—has been widely explored in recent years \cite{laptev2020you}\cite{zheng2021using}\cite{casanova2022asr}. These models are becoming increasingly capable of producing natural, high-quality, and expressive synthetic speech that closely resembles human recordings \cite{kong2020hifi}. Consequently, synthetic speech has become an attractive alternative for supplementing real data in various speech processing tasks.

Various studies have investigated the effectiveness of synthetic data for fine-tuning ASR systems especially for English. Tomoya Mizumoto et al. \cite{mizumoto2025synthetic} demonstrate that training solely on synthetic data can degrade performance, whereas combining synthetic data with real speech leads to performance improvements, highlighting its potential when used alongside real data. Benedikt Hilmes et al. \cite{hilmes2024effect} analyze how the performance gap between models trained on synthetic versus real data varies with changes in speaker embeddings and model scaling.

The primary objective of this study is to evaluate the effectiveness of synthetic data in improving Automatic Speech Recognition(ASR) Systems. Specifically, we aim to analyze how the inclusion of synthetic data impacts model performance and whether it can serve as a viable substitute or supplement to real data. Our experiments focus on three Indic languages: Hindi, Kannada, and Telugu. These languages were selected as they represent two major language families in India—Indo-Aryan and Dravidian—thus providing a broader perspective on the applicability of synthetic data across different linguistic groups. 

In this study, We first quantify the performance gains obtained by augmenting synthetic data with  real speech data, analyzing how the addition of synthetic data with real data influences ASR performance. We then independently examine how ASR accuracy varies with different sources of textual scripts used for synthetic speech generation, highlighting the role of linguistic diversity and script quality in model training. Furthermore, we evaluate the effectiveness of synthetic speech generated using multiple speech synthesis models to understand how the output affect downstream ASR performance. Finally, we study the impact of voice cloning in synthetic speech generation by analyzing how ASR performance changes with the number and diversity of distinct cloned voices used during data generation, providing insights into speaker variability requirements for effective synthetic data augmentation.
\section{Synthetic Data Generation}
\begin{table*}[t]
\caption{Performance comparison across Hindi, Kannada, and Telugu datasets under different fine-tuning strategies}
\centering
\small
\setlength{\tabcolsep}{9pt}
\begin{tabular}{llcccccc}
\toprule
\textbf{Language} & \textbf{Dataset} &
\makecell{\textbf{Without}\\\textbf{FT}} &
\textbf{Vaani} &
\makecell{\textbf{Vaani +}\\\textbf{Syn.}} &
\makecell{\textbf{Vaani +}\\\textbf{RESPIN}} &
\makecell{\textbf{Only}\\\textbf{RESPIN}} &
\makecell{\textbf{Only}\\\textbf{Syn.}} \\
\midrule

\multirow{9}{*}{\textbf{Hindi}}
& GramVaani        & 1.6744 & 0.5007 & 0.4640 & 0.4630 & 0.5781 & 0.6552 \\
& FLEURS           & 0.8293 & 0.3036 & 0.2450 & 0.2390 & 0.2674 & 0.3053 \\
& MUCS             & 1.5935 & 0.3744 & 0.3306 & 0.3234 & 0.3766 & 0.4928 \\
& CommonVoice      & 1.1093 & 0.3905 & 0.3489 & 0.3370 & 0.4009 & 0.5329 \\
& Kathbath         & 0.8768 & 0.2734 & 0.2199 & 0.2143 & 0.2588 & 0.3118 \\
& Kathbath\_noisy  & 0.9275 & 0.3050 & 0.2498 & 0.2389 & 0.2861 & 0.3500 \\
& Vaani            & 1.8684 & 0.2670 & 0.2583 & 0.2611 & 0.4636 & 0.5787 \\
& RESPIN           & 1.2393 & 0.2723 & 0.1658 & 0.1304 & 0.1191 & 0.1916 \\
\cmidrule(lr){2-8}

& \textbf{AVG}     & \textbf{1.2648} & \textbf{0.3358} & \textbf{0.2852} & \textbf{0.2758} & \textbf{0.3438} & \textbf{0.4272} \\
\midrule

\multirow{6}{*}{\textbf{Kannada}}
& FLEURS           & 1.0934 & 0.4769 & 0.4026 & 0.3500 & 0.3964 & 0.5829 \\
& Kathbath         & 1.2432 & 0.5225 & 0.4665 & 0.4012 & 0.4565 & 0.6749 \\
& Kathbath\_noisy  & 1.2442 & 0.5667 & 0.5117 & 0.4474 & 0.4999 & 0.7209 \\
& Vaani            & 1.6277 & 0.7237 & 0.7259 & 0.7242 & 0.8606 & 0.9626 \\
& RESPIN           & 1.4954 & 0.5089 & 0.3884 & 0.2814 & 0.2854 & 0.4853 \\
\cmidrule(lr){2-8}
& \textbf{AVG}     & \textbf{1.3407} & \textbf{0.5500} & \textbf{0.4990} & \textbf{0.4408} & \textbf{0.4997} & \textbf{0.6853} \\
\midrule

\multirow{6}{*}{\textbf{Telugu}}
& FLEURS           & 1.5818 & 0.5483 & 0.4701 & 0.3941 & 0.4440 & 0.6015 \\
& Kathbath         & 1.4493 & 0.5633 & 0.4878 & 0.4158 & 0.4716 & 0.6362 \\
& Kathbath\_noisy  & 1.5085 & 0.6085 & 0.5315 & 0.4528 & 0.5113 & 0.6825 \\
& Vaani            & 2.6605 & 0.5428 & 0.5357 & 0.5162 & 0.7168 & 0.8451 \\
& RESPIN           & 2.0411 & 0.4780 & 0.3541 & 0.2558 & 0.2761 & 0.4453 \\
\cmidrule(lr){2-8}
& \textbf{AVG}     & \textbf{1.8482}    & \textbf{0.5481} & \textbf{0.4758} & \textbf{0.4069} & \textbf{0.4840} & \textbf{0.6401} \\
\bottomrule
\end{tabular}

\label{tab:finetuning_results_all}
\end{table*}
To generate synthetic speech data, we employed Text-to-Speech (TTS) models with voice cloning capabilities. Specifically, we used the Coqui TTS framework, based on the VITS architecture proposed by Kim et al.~\cite{kim2021vits}, which enables high-quality end-to-end neural speech synthesis with multilingual support. Hindi was natively supported within the framework. For Telugu and Kannada, which were not natively supported in our setup, we fine-tuned the base VITS model using the SYSPIN dataset~\cite{syspin_s1.0_2025} to adapt it to the target languages. The SYSPIN dataset provides high-quality studio-recorded speech data across nine Indian languages, including approximately 101 hours for Kannada and 105 hours for Telugu. The availability of clean, well-annotated recordings makes it particularly suitable for fine-tuning neural TTS systems for synthetic data generation.

To investigate the impact of script source on ASR training, we curated textual prompts from a range of sources, including well-established Indic speech corpora such as RESPIN\cite{kumarRESPIN}, IndicVoices\cite{javed2024indicvoices}, and Kathbath\cite{javed2023indicsuperb}. Additionally, we leveraged large language models (LLMs) like Gemini 2.5 Flash Lite\cite{comanici2025gemini} and Gemini 3 Flash\cite{pichai2025gemini3} to generate scripts, allowing us to assess how LLM-generated text affects downstream ASR performance. The same set of prompts was used across the various models to generate the descriptions. For the generation process, we used Coqui-XTTS-v2 and a speaker from the SYSPIN Hindi subset as the reference audio.

To analyze the impact of the TTS model on ASR performance, we compared three models: Coqui-XTTS-v2, IndicParlor TTS\cite{sankar25_interspeech}\cite{lacombe-etal-2024-parler-tts}\cite{lyth2024natural}, and IndriTTS\cite{indri-multimodal-alm}. For voice synthesis, we utilized Gemini 2.5 Flash.  For  Coqui-XTTS-v2, the reference audio was taken from a  speaker in the SYSPIN Hindi subset, while for IndicParlor TTS, we used Gemini 2.5 Flash Lite to generate different speaker descriptions.

\section{Experiments}

We evaluated performance gains using WER with the Whisper Small ASR model. For all fine-tuning experiments, the Whisper models were trained using a learning rate of $1 \times 10^{-5}$ with 1,000 warmup steps. 
Training was conducted for up to 20 epochs with a per-device batch size of 32, utilizing FP16 mixed precision to optimize memory and compute efficiency.
Model checkpoints were evaluated and saved every 500 steps, and the final model was selected based on the lowest Word Error Rate (WER) achieved on the validation set.  Experiments were conducted across three languages—Hindi, Kannada, and Telugu—to assess the generalizability of the approach. The spontaneous data from the Vaani dataset\cite{Vaani2025} was augmented with synthetic data. The model was fine-tuned under multiple data configurations: (i) Vaani data only, (ii) Synthetic data generated exclusively using RESPIN transcriptions, (iii) Vaani combined with original RESPIN data, (iv) RESPIN data only,  and (v) Vaani augmented with synthetic data generated from RESPIN transcriptions. This experimental setup allows for a systematic analysis of the impact of synthetic data on ASR performance.

\begin{table*}[t]
\caption{Effect of Gemini-generated synthetic data scale on ASR performance}
\centering
\small
\setlength{\tabcolsep}{6pt}
\begin{tabular}{lcccccc}
\toprule
\textbf{Dataset} &
\makecell{\textbf{Without}\\\textbf{Finetuning}} &
\textbf{\#Speakers 1} &
\textbf{\#Speakers 10} &
\textbf{\#Speakers 100} &
\textbf{\#Speakers 1000} &
\textbf{\#Speakers 10000} \\
\midrule

GramVaani        & 1.6744 & 0.8176 & 0.7613 & 0.7940 & 0.8055 & 0.7793 \\
FLEURS           & 0.8293 & 0.3945 & 0.3512 & 0.3552 & 0.3447 & 0.3528 \\
MUCS             & 1.5935 & 0.7210 & 0.6403 & 0.7445 & 0.7166 & 0.6896 \\
CommonVoice      & 1.1093 & 0.7896 & 0.6565 & 1.0441 & 1.0263 & 1.0564 \\
Kathbath         & 0.8768 & 0.4574 & 0.4038 & 0.5849 & 0.5077 & 0.5917 \\
Kathbath\_noisy  & 0.9275 & 0.5080 & 0.4447 & 0.5340 & 0.5109 & 0.5409 \\
Vaani            & 1.8684 & 0.7315 & 0.6798 & 0.7535 & 0.7548 & 0.7441 \\
RESPIN           & 1.2393 & 0.4854 & 0.4266 & 0.5290 & 0.4935 & 0.5176 \\
\midrule
\textbf{Avg}     & \textbf{1.2648} & \textbf{0.6131} & \textbf{0.5455} & \textbf{0.6674} & \textbf{0.6450} & \textbf{0.6591} \\
\bottomrule
\end{tabular}

\label{tab:gemini_scale_results}
\end{table*}
For Hindi, we used 532.9257 hours (473.8215 hours for training and 59.1042 hours for validation) of real speech data from the Vaani corpus, along with 130.68 hours (train: 128.47 hours, validation: 2.21 hours) of data from the RESPIN dataset. Additionally, we generated 164.6658 hours (train: 161.7282 hours, validation: 2.9376 hours) of synthetic speech using RESPIN transcriptions. The synthetic audio was generated using the Coqui TTS framework with voice cloning enabled. For voice selection, each utterance was synthesized by randomly sampling a speaker from a pool of 10,000 speakers available in the Vaani dataset, ensuring speaker diversity. We followed the same data generation protocol for Telugu and Kannada. Specifically, for Telugu, we used 102.2679 hours (train: 92.1480 hours, validation: 10.1199 hours) of real speech data from Vaani, and for Kannada, we used 108.1747 hours (train: 96.8835 hours, validation: 11.2912 hours) of real speech data from Vaani. For the RESPIN dataset, we used 158.1912 hours (train: 155.8891 hours, validation: 2.3021 hours) of Telugu data and 167.20 hours (train: 164.8344 hours, validation: 2.3659 hours) of Kannada data. Synthetic data for both languages was generated using RESPIN transcriptions and the same speaker sampling strategy used for Hindi. The synthetic data generated accounts for 186.6314 hours for Kannada (train: 183.77 hours, validation: 2.8581 hours) and 208.3129 hours for Telugu (train: 197.7489 hours, validation: 10.5640 hours).

For evaluation, we employed multiple benchmark datasets to ensure a comprehensive and reliable assessment of model performance across diverse recording conditions, collection methodologies, and speaker demographics. This multi-dataset evaluation strategy allows for a robust analysis of model generalization in realistic deployment scenarios. The Hindi models were evaluated on GramVaani\cite{bhanushali2022gram}, FLEURS\cite{conneau2023FLEURS}, MUCS\cite{Diwan_2021}, Common Voice\cite{ardila2020common}, Kathbath (including its noisy variant), Vaani, and RESPIN. The Telugu models were evaluated on FLEURS, Kathbath (including its noisy variant), Vaani, and RESPIN. The Kannada models were evaluated on FLEURS, Kathbath (including its noisy variant), and Vaani.

To analyze the impact of the number of speakers used during the voice cloning process, we generated synthetic Hindi speech using transcriptions produced by Gemini 2.5 Flash Lite and speech synthesis via the Coqui-XTTS-v2 framework. All experiments were conducted on the Hindi subset. We generated around 43 hours (train: around 32 hours, validation: around 11 hours) of synthetic speech under varying speaker diversity settings by randomly sampling reference speakers from the Vaani dataset. Specifically, we synthesized data using 1, 10, 100, 1,000, and 10,000 distinct speakers. Separate ASR models were fine-tuned for each configuration and evaluated independently to quantify the effect of speaker diversity in synthetic data generation.

To analyze the impact of the transcription source on the generation of synthetic data, we conducted controlled experiments exclusively in Hindi. Speech was synthesized using the Coqui-XTTS-v2 framework with voice cloning enabled. For consistency, we used a single reference speaker from the SYSPIN data set for all synthetic utterances. Transcriptions were sourced from multiple origins to isolate script effects. These included manually curated read-speech corpora such as RESPIN, IndicVoices, and Kathbath. In addition, we generated scripts using large language models, specifically Gemini 2.5 Flash Lite and Gemini 3 Flash, to compare human-curated and LLM-generated transcription sources. For each transcription source, we generated 10 hours of synthetic speech for training and 1 hour for validation.

To study the impact of model size on performance gains, we fine-tuned multiple variants of the Whisper architecture, including Tiny, Base, Small, Medium, and Large. All models were trained using synthetic Hindi data generated from transcriptions produced by Gemini 2.5 Flash lite. Speech synthesis was performed using voice cloning with a single reference speaker from the SYSPIN dataset to maintain consistency across experiments. Each Whisper variant was fine-tuned independently and evaluated across multiple Hindi benchmark datasets to assess how model capacity influences performance gains from synthetic data.

\section{Results}

 As shown in Table \ref{tab:finetuning_results_all}, the  fine-tuned models consistently show performance improvements when augmented with synthetically generated data. For models trained on Vaani data, adding synthetic speech generated using RESPIN transcriptions reduces the average  Word Error Rate (WER) by 15.06\% for Hindi, 9.27\% for Kannada, and 13.19\% for Telugu. Although the gains achieved through synthetic data augmentation are slightly lower than those obtained by incorporating additional real speech data, they remain substantial across all languages. Notably, training with synthetic data alone also leads to measurable improvements, reducing WER by 62.68\% for Hindi and 48.85\% for Kannada and  65.21\% for Telugu. These results indicate that synthetically generated speech can serve as an effective substitute for real data in low-resource settings, while also providing complementary gains when used alongside real speech corpora.
\begin{figure}[t]
    \centering
    \includegraphics[width=\linewidth]{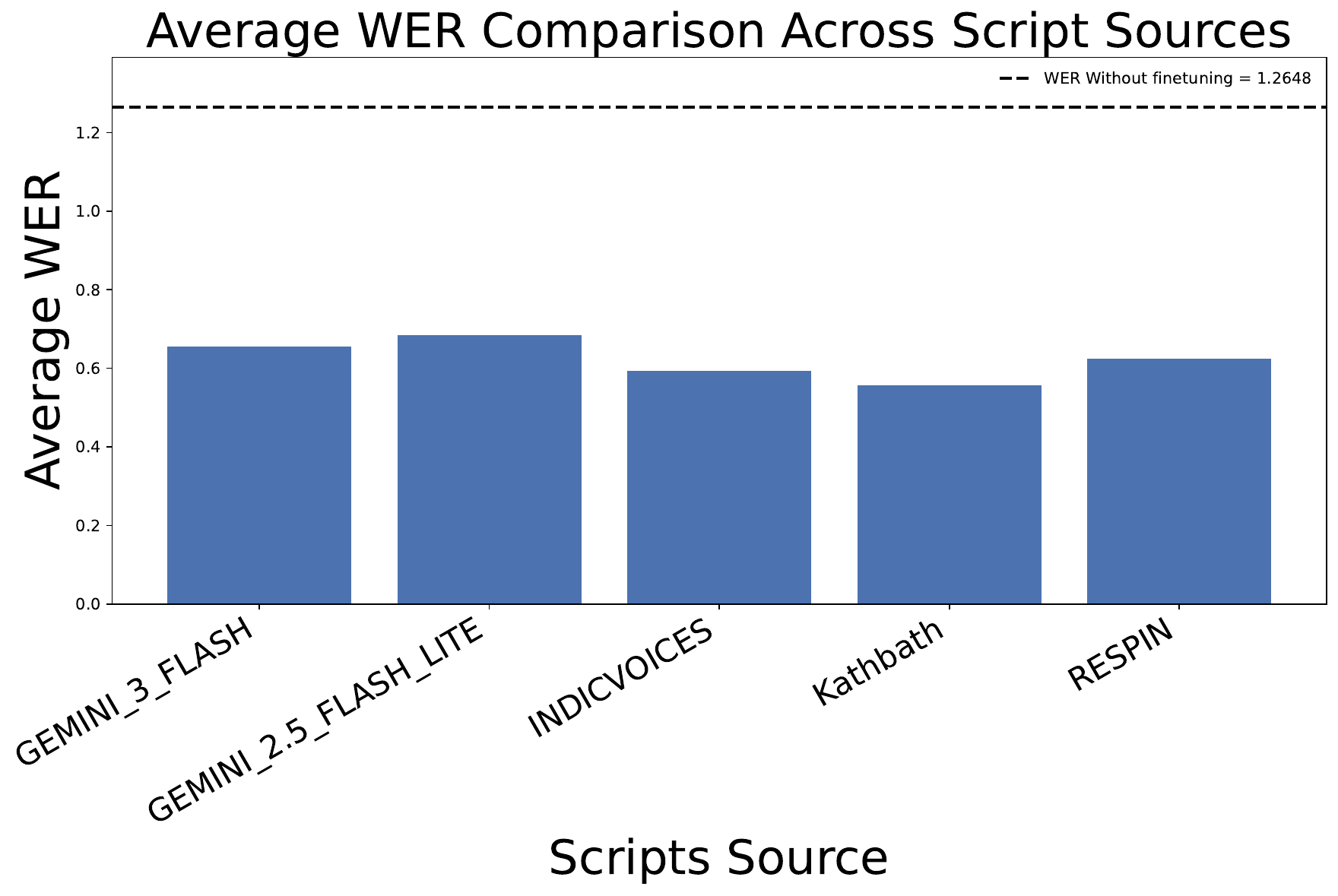}
    \caption{Average WER comparison across script sources. The dashed horizontal line indicates the baseline WER (1.2648).}
    \label{fig:avg_wer_script_sources}
\end{figure}

From the performance of models trained with varying numbers of speakers in the voice cloning process, as shown in  table \ref{tab:gemini_scale_results} we observe a decrease in WER when increasing the number of speakers from 1 to 10. Beyond this point, no further reduction in WER is observed. This behavior suggests that limited speaker diversity is sufficient to capture the benefits of voice variation in synthetic data for the model under evaluation. Increasing the number of speakers beyond this threshold does not yield additional gains, likely because the ASR models have already been exposed to substantial speaker diversity during their initial pre-training. As a result, further increases in speaker variation during fine-tuning provide diminishing returns.

\begin{figure}[t]
    \centering
    \includegraphics[width=\linewidth]{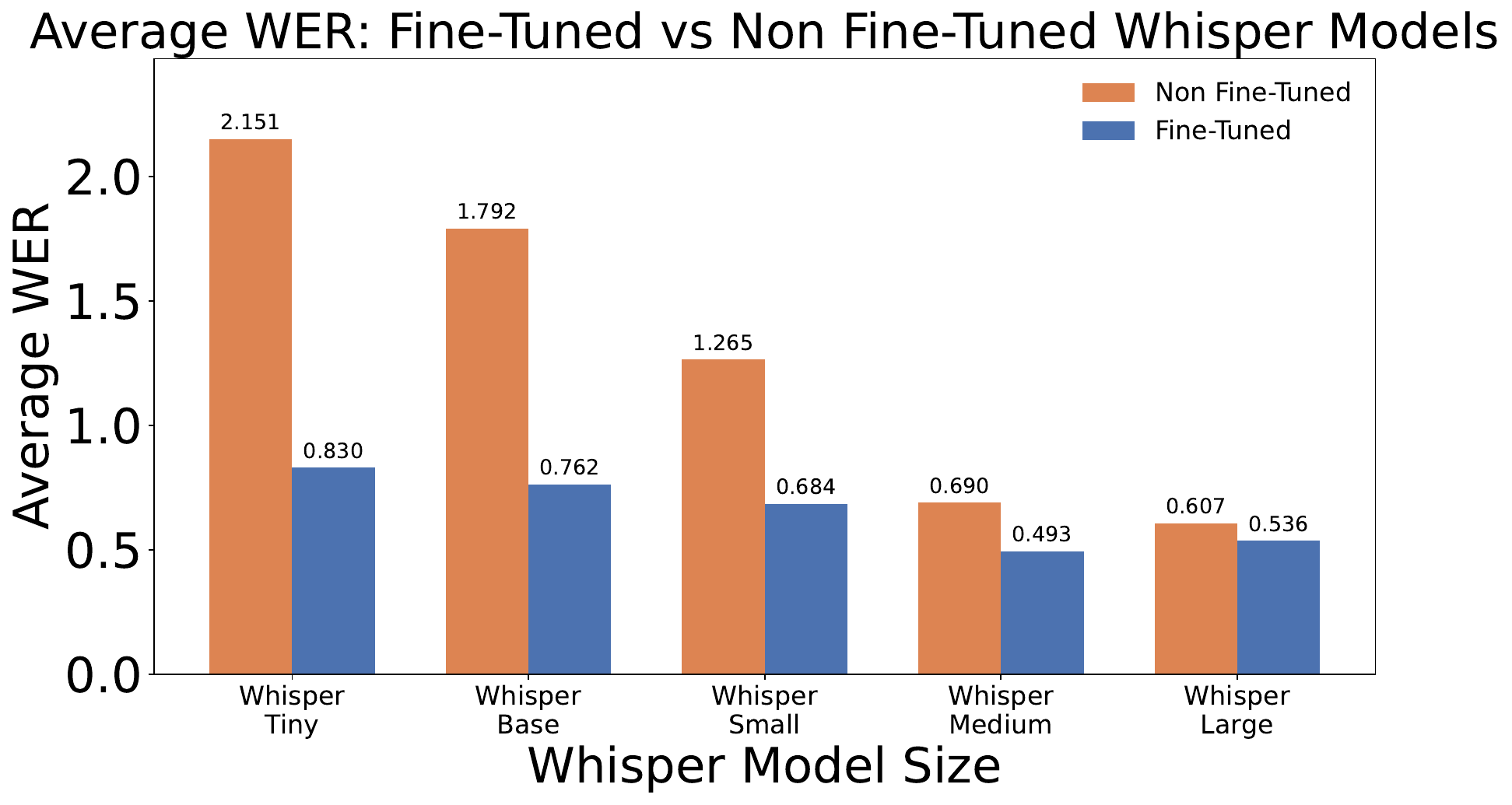}
    \caption{Average WER comparison across different whisper models}
    \label{fig:avg_wer_whisper_models}
\end{figure}

The evaluation across different script sources demonstrates consistent performance improvements across all models when trained using data from each source, as shown in figure \ref{fig:avg_wer_script_sources}. While gains are observed in all cases, the magnitude of improvement varies slightly depending on the transcription source, indicating that transcription quality and style influence model performance. Furthermore, different versions of the Whisper models exhibit varying degrees of performance gains, as shown in figure \ref{fig:avg_wer_whisper_models}, suggesting that model scale and architecture interact differently with the characteristics of the transcription data.

\begin{table}[th]
  \caption{Word Error Rate (WER) of Whisper Small (baseline) versus models finetuned on various TTS datasets}
  \label{tab:whisper_tts_finetune_wer}
  \centering
  \resizebox{\columnwidth}{!}{%
  \begin{tabular}{l c ccc}
    \toprule
    \multirow{2}{*}{\textbf{Dataset}} & \textbf{Whisper} & \multicolumn{3}{c}{\textbf{Finetuned on TTS Dataset}} \\
    \cmidrule(lr){3-5}
    & \textbf{(No FT)} & \textbf{IndriTTS} & \textbf{CoquiXTTSv2} & \textbf{Indic Parlor TTS} \\
    \midrule
    Gramvaani      & 1.674 & \textbf{0.860} & 1.085 & 0.952 \\
    Fleurs         & 0.829 & 0.413 & 0.447 & \textbf{0.396} \\
    MUCS           & 1.594 & \textbf{0.678} & 0.763 & 0.822 \\
    CommonVoice    & 1.109 & 0.634 & \textbf{0.604} & 0.868 \\
    Kathbath       & 0.877 & \textbf{0.461} & 0.493 & 0.481 \\
    Kathbath Noisy & 0.928 & 0.496 & 0.524 & \textbf{0.495} \\
    Vaani          & 1.868 & \textbf{0.763} & 0.998 & 0.971 \\
    Respin         & 1.239 & 0.506 & 0.557 & \textbf{0.493} \\
    \midrule
    \textbf{Average WER} & 1.265 & \textbf{0.601} & 0.684 & 0.685 \\
    \bottomrule
  \end{tabular}%
  }
\end{table}

The performance gain obtained using data generated from different TTS sources (Table \ref{tab:whisper_tts_finetune_wer}) shows slight variation across the models. Among them, IndriTTS provides the highest performance improvement with a gain of 52.49\%, followed by CoquiXTTSv2 with 45.93\% and Indic Parlor with 45.85\%.
\section{Conclusion}
From our experiments, it is evident that incorporating synthetic data leads to consistent improvements in ASR performance across all evaluated languages. However, the observed performance gains remain lower than those achieved using real-world data, highlighting the continued importance of authentic recordings. Additionally, the source and quality of the textual data used to generate synthetic speech significantly influence model performance. Notably, synthetic speech generated through voice cloning yields stronger improvements than standard text-to-speech. Increasing the number of speakers from 1 to 10 reduces WER, but beyond this point no further gains are observed for heavily pre-trained models, indicating that moderate speaker diversity is sufficient and additional speakers provide diminishing returns.
\bibliographystyle{IEEEtran}
\bibliography{mybib}

\end{document}